\documentclass[11pt]{extarticle}

\usepackage[T1]{fontenc}
\usepackage[utf8]{inputenc} % safe for pdfLaTeX; harmless on modern LaTeX
\usepackage{lmodern}
\usepackage{microtype}

\usepackage[
  letterpaper,
  left=1in,
  right=1in,
  top=1in,
  bottom=1in,
  includeheadfoot
]{geometry}

\usepackage{setspace}
\usepackage{amsmath,amssymb,bm}
\usepackage{graphicx}
\usepackage{caption}
\usepackage{subcaption}

\usepackage{authblk}

\usepackage{xcolor}
\usepackage{hyperref}
\hypersetup{
  colorlinks=true,
  linkcolor=blue,
  citecolor=blue,
  urlcolor=blue,
  pdfauthor={Ingrid Torres; Alex Krasnok},
  pdftitle={Rotational Tuning, Moir\'e-like Field Patterns, and Strong Coupling in Double-Layer Wire Metasurfaces}
}

\title{\textbf{Twist-Tuned Strong Coupling in Sub-GHz Wire Metasurface Bilayers}}
\author[1]{Ingrid Torres}
\author[1,2]{Alex Krasnok\thanks{To whom correspondence should be addressed: \texttt{akrasnok@fiu.edu}}}
\affil[1]{Department of Electrical and Computer Engineering, Florida International University, Miami, Florida 33174, USA}
\affil[2]{Knight Foundation School of Computing and Information Sciences, Florida International University, Miami, Florida 33199, USA}
\date{December 2025}

\begin{document}
\maketitle

\begin{abstract}
Twist-angle control offers a bias-free route to reconfigurable metasurfaces, yet its extension to deeply subwavelength resonant platforms at VHF/UHF remains limited. We demonstrate a sub-GHz double-layer wire metasurface formed by two identical wire grids separated by a gap $G$, with in-plane rotation $\phi$ as the sole tuning parameter. One-port, loop-coupled $S_{11}$ measurements supported by full-wave simulations reveal twist-driven hybridization of the dominant resonant manifold. For small $G$, the lower hybrid resonance redshifts continuously from $\sim409$~MHz to $\sim210$~MHz ($>2{:}1$ tuning), enabling compact, twist-programmable resonant surfaces. Simulations further show that twisting imprints moir\'e-like magnetic near-field super-modulations. From resonance frequencies, linewidths, and normal-mode splitting extracted from the complex response, we obtain normalized coupling up to $g\approx0.43$ with cooperativity exceeding unity over broad angular ranges, meeting the resolved-splitting criterion. The rapid collapse of tunability at larger $G$ confirms the near-field origin of the interaction.
\end{abstract}

%========================
% Introduction (revised)
%========================
\section{Introduction}
The discovery that a relative twist between stacked two-dimensional materials can reshape electronic bands and correlations has catalyzed a broader search for twist-controlled analogues in classical wave physics \cite{ref1,ref2,ref3,ref6,ref8,ref10}. In electromagnetics and photonics, the twist angle between anisotropic patterned sheets modifies their mutual electromagnetic coupling and thereby tunes polarization conversion, chirality, and dispersion in layered metasurfaces \cite{ref12,ref13,ref14,ref15,liu2022magic,orazbay2024polar}. Rotation has also emerged as a sensitive control variable in nonlinear metasurfaces, where small angular changes can switch multistable states \cite{valagiannopoulos2025nonlinear,dellavalle2017nonlinear}. Despite this rapid progress, an important gap remains for the antennas-and-propagation community: experimentally grounded twist control of \emph{resonant} metasurface platforms in the \emph{sub-GHz} regime, where near-field interactions are exceptionally strong, device footprints are intrinsically large, and common active-tuning approaches (bias networks, varactors, switches) often impose loss, complexity, and power-handling constraints. Establishing twist as a practical, quantitative knob in this frequency range requires not only demonstration of large, continuous tuning, but also a clear pathway to extracting coupling and loss metrics from standard microwave measurements.

Wire-based metasurfaces provide a natural and physically transparent testbed for this objective. Wire grids underpin classic frequency-selective surfaces and strongly anisotropic artificial media \cite{ref29,ref30,ref33}, and when two such grids are brought into close proximity their interaction is dominated by near fields that are highly sensitive to relative alignment. A mutual rotation between the layers can therefore act as a purely geometric control parameter that continuously reshapes interlayer coupling without changing materials, adding lumped components, or introducing bias circuitry. At the same time, twisting two periodic wire patterns produces moir\'e-like spatial beating, so that the relevant electromagnetic near-field texture need not be set by geometry alone, but can instead reflect the hybridized response of the coupled bilayer. The opportunity, then, is to use twist to co-program \emph{both} the resonant spectrum and the magnetic near-field distribution in a deeply subwavelength, strongly coupled platform at VHF/UHF.

In this work we investigate a double-layer wire metasurface (DLWM) formed by two identical wire arrays separated by a low-permittivity spacer of thickness $G$ and rotated by an in-plane angle $\phi$. The structure operates in the sub-GHz band in a macroscale prototype that allows repeatable control of $G$ and $\phi$ while remaining fully compatible with Maxwell scaling to printed and higher-frequency implementations. We characterize the bilayer using a compact one-port magnetic-loop excitation and measure $S_{11}(f,\phi,G)$. Because loop-driven measurements can exhibit probe- and environment-sensitive line-shape distortions, we base all quantitative conclusions on resonance frequencies, linewidths, and splittings extracted directly from the complex response using robust fitting procedures \cite{probst2015robust,khalil2012analysis}. Full-wave simulations are used to link the observed spectral evolution to the underlying current/field distributions and to visualize the twist-programmable magnetic near-field textures. Together, these results establish twist as a bias-free mechanical control knob for sub-GHz resonant metasurfaces, and provide a clear experimental pathway to quantifying how interlayer coupling evolves with rotation and separation. 

%%%%%%%%%%%%%%%%

%========================
% Results
%========================
\section{Results}

\begin{figure}[!t]
\centering
\includegraphics[width=0.7\linewidth]{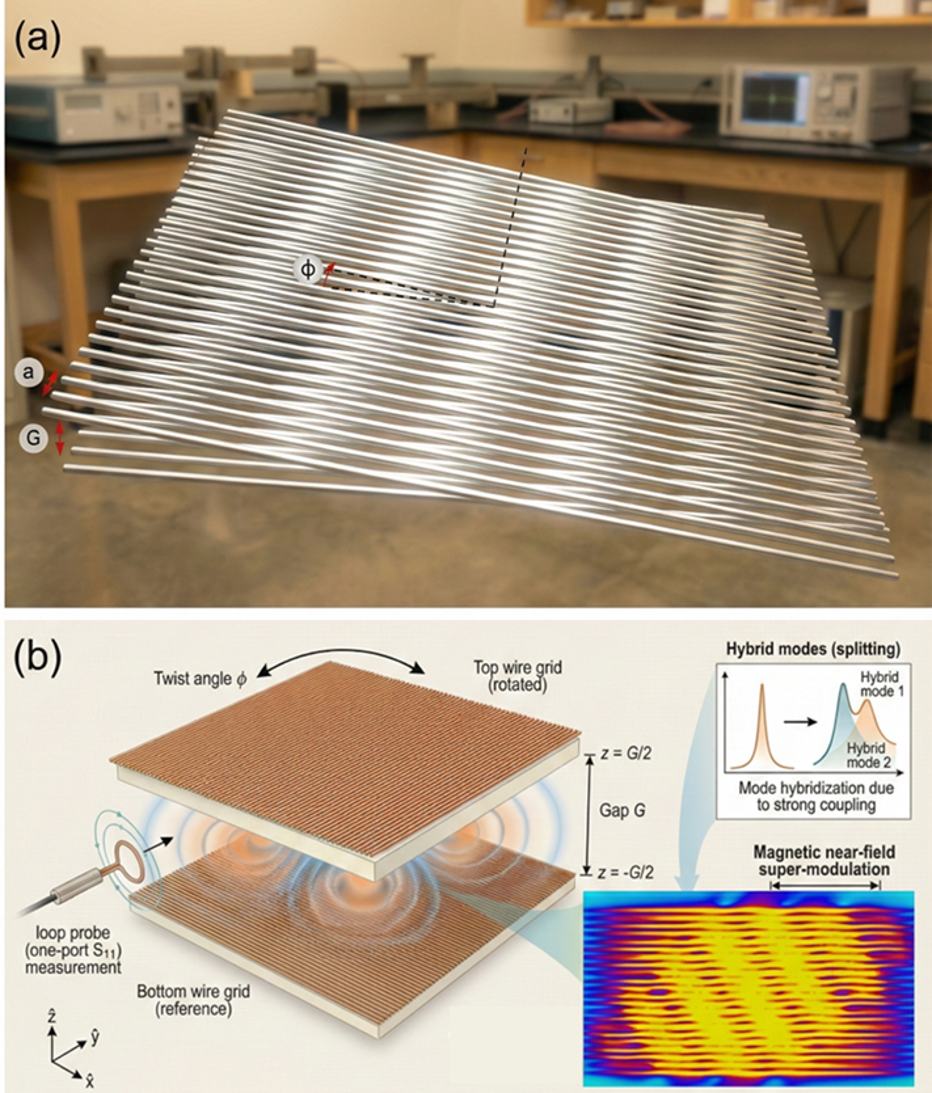}
\caption{Twist-controlled double-layer wire metasurface (DLWM) platform and operating concept. (a) Sketch of the macroscale prototype showing two identical copper-wire grids (each with $N_w=31$ wires of length $L=350~\mathrm{mm}$, radius $r=0.9~\mathrm{mm}$, and in-plane period $a\approx5~\mathrm{mm}$) separated by a low-permittivity foam spacer of thickness $G$ ($\varepsilon_r\approx1.05$). The top grid is rotated in-plane by the twist angle $\phi$ relative to the bottom grid. (b) Schematic of the measurement and the two key twist-enabled effects explored in this work: twist-dependent hybridization (mode splitting/dispersion in the one-port $S_{11}$ response) and twist-programmable magnetic near-field super-modulation. The two grids lie in planes $z=\pm G/2$. For a fixed spacer thickness $G$, the only in-situ tuning knob is the relative rotation $\phi$; varying $G$ is used to access weak- and strong-coupling regimes. We exploit the symmetry $S_{11}(f,\phi)=S_{11}(f,-\phi)=S_{11}(f,\phi+\pi)$ and therefore restrict $\phi\in[0,\pi/2]$.}
\label{fig:dlwm_schematic}
\end{figure}

Figure~\ref{fig:dlwm_schematic}(a) shows the DLWM platform studied in this work. Two identical finite wire grids are brought into close proximity and tuned only by the in-plane twist angle $\phi$ at fixed interlayer separation $G$. This configuration targets a regime that is simultaneously attractive and challenging for antennas and propagation: VHF/UHF resonant surfaces are naturally large, and common electronic tuning strategies often introduce loss and biasing complexity, whereas purely geometric twist provides a passive, reversible control knob whose action is mediated by near fields. Figure~\ref{fig:dlwm_schematic}(b) summarizes the experimental configuration and the two observables that motivate the analysis that follows: twist-controlled hybridization of the dominant resonant response (tracked via the one-port $S_{11}$ spectrum) and twist-programmable magnetic near-field super-modulation (quantified from full-wave field maps). In the small-$G$ regime, rotation predominantly reshapes the mutual near-field interaction between the two grids, enabling large, continuous tuning without modifying materials or adding bias networks.

The theoretical approach used throughout the paper is intentionally minimal and is constructed to interface directly with driven spectra and full-wave field maps. Each wire supports a fundamental standing-wave current distribution along its axis, and the dominant resonance of a finite wire array can be viewed as a collective current pattern concentrated along the wires. For a wire of length $L$, an approximate first-mode current profile may be written as
\begin{equation}
I(u,t)=I_0\,\sin\!\big[k_0\big(\tfrac{L}{2}-|u|\big)\big]\cos(\omega_0 t),
\label{eq:wire_current}
\end{equation}
with $k_0\approx\pi/L$, emphasizing magnetic-energy concentration near midpoints (current antinode) and electric-energy concentration near ends (charge accumulation). Because the arrays are finite and the excitation/readout is local (Fig.~\ref{fig:dlwm_schematic}(b)), additional weak features can appear due to edge perturbations and probe/environment loading; therefore, all quantitative conclusions in later sections are based on resonance frequencies, linewidths, and splittings extracted from the complex response rather than on fine line-shape details.

At the sub-GHz operating frequencies of interest, $a\ll\lambda$, and each grid can be represented in a homogenized sense as an anisotropic impedance sheet (or, more generally, within a generalized sheet-transition-condition framework),
\begin{equation}
\bm{E}_\parallel(x,y)=\overline{\overline{Z}}_s(\omega)\cdot \bm{J}_\parallel(x,y),\qquad
\overline{\overline{Z}}_s(\omega)=
\begin{pmatrix}
Z_{xx}(\omega) & 0\\[2pt]
0 & Z_{yy}(\omega)
\end{pmatrix},
\label{eq:Zsheet}
\end{equation}
where the wire-parallel response is resonant and may be approximated by a Lorentz-type form,
\begin{equation}
Z_{xx}(\omega)\approx R_{s,\mathrm{eff}}+j\!\left(\omega L_{s,\mathrm{eff}}-\frac{1}{\omega C_{s,\mathrm{eff}}}\right),
\label{eq:lorentzZ}
\end{equation}
while the transverse response is predominantly capacitive, $Z_{yy}(\omega)\approx 1/(j\omega C_{s,\mathrm{grid}})$ \cite{ref40,ref41,munk2000fss,tretyakov2003analytical,holloway2012metasurfaces,kuester2003metafilm}. Rotation of the top layer by $\phi$ is captured by the standard tensor rotation
\begin{equation}
\overline{\overline{Z}}_s^{(2)}(\phi)=\mathbf{R}(\phi)\,\overline{\overline{Z}}_s^{(1)}\,\mathbf{R}^{T}(\phi),
\qquad
\mathbf{R}(\phi)=
\begin{pmatrix}
\cos\phi & -\sin\phi\\
\sin\phi & \cos\phi
\end{pmatrix},
\label{eq:rotation_tensor}
\end{equation}
which yields off-diagonal components in the laboratory frame when $Z_{xx}\neq Z_{yy}$. Geometric overlay of two identical 1D lattices yields the moir\'e length
\begin{equation}
\Lambda_{\mathrm{geo}}=\frac{a}{2|\sin(\phi/2)|},
\label{eq:Lambda_geo}
\end{equation}
but the experimentally relevant super-modulation scale is electromagnetic: it reflects the dominant hybridized near-field content rather than geometry alone. In practice, we quantify the electromagnetic superperiod $\Lambda_{\mathrm{EM}}$ from magnetic-field maps by measuring the dominant fringe spacing (e.g., by line cuts normal to the modulation direction or by the first off-center maximum of the 2D autocorrelation of $F_H$), and we use $\Lambda_{\mathrm{geo}}$ only as a geometric reference scale.

\begin{figure}[!t]
\centering
\includegraphics[width=0.98\linewidth]{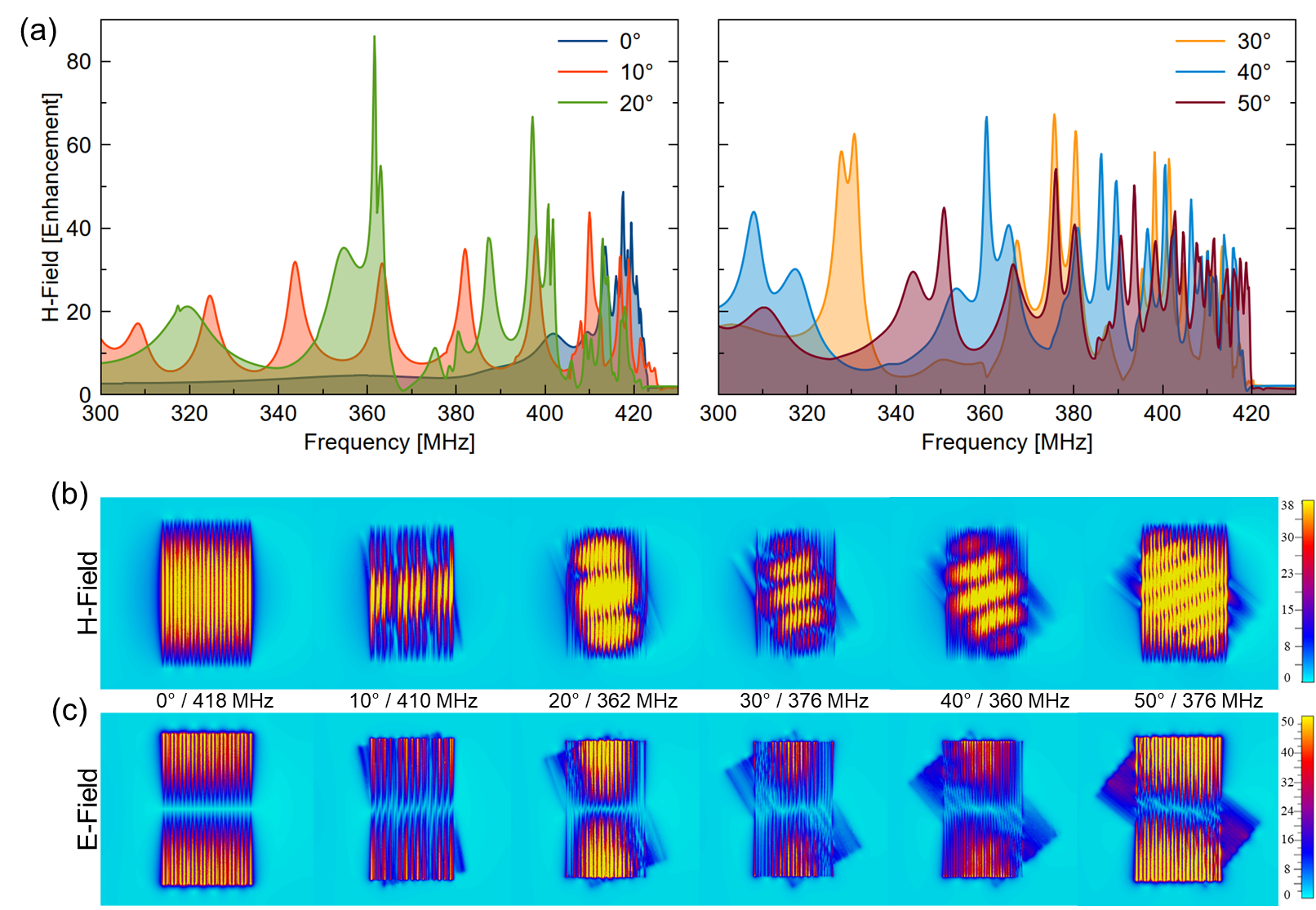}
\caption{Plane-wave simulation of twist-dependent hybridization and near fields for $G=2$~mm. (a) Peak magnetic-field enhancement at the mid-gap plane $z=0$ versus frequency for several twist angles. (b) Representative $F_H$ maps and (c) $F_E$ maps at selected resonant frequencies. Twist induces hybridized resonances and programmable moir\'e-like magnetic super-modulations; in the text, the dominant superperiod $\Lambda_{\mathrm{EM}}$ is obtained directly from the fringe spacing of $F_H$ (e.g., from line cuts or autocorrelation) and compared to the geometric scale $\Lambda_{\mathrm{geo}}$.}
\label{fig:rotation_spectra_fields}
\end{figure}

To visualize twist-programmable field textures, we use plane-wave excitation (normal incidence, $\bm{E}_{\mathrm{inc}}\parallel\hat{x}$) in full-wave simulations and report the enhancement for any field component $X\in\{H,E\}$ as
\begin{equation}
F_X(\bm{r};f,\phi)=\frac{|X(\bm{r};f,\phi)|}{|X_{\mathrm{inc}}|}.
\label{eq:enhancement}
\end{equation}
Figure~\ref{fig:rotation_spectra_fields} summarizes representative simulated spectra and field maps used later to connect spectral evolution to spatial reconfiguration.

The spectral response of the bilayer is modeled by retaining a single dominant resonant degree of freedom for each grid and coupling the two through both mutual inductive and capacitive interactions. In a minimal LC/coupled-resonator description, each layer is represented as an $(L,C)$ resonator coupled through mutual inductance $M(\phi,G)$ and mutual capacitance $C_m(\phi,G)$, yielding modal frequencies (for details, see \textit{SM Sec.~S1})
\begin{equation}
\omega_{\pm}^{2}=\omega_{0}^{2}\,\frac{1\pm k_m(\phi,G)}{1\pm k_e(\phi,G)},
\qquad
\omega_0=\frac{1}{\sqrt{LC}},
\label{eq:omega_pm}
\end{equation}
where the dimensionless coupling coefficients are
\begin{equation}
k_m(\phi,G)=\frac{M(\phi,G)}{L},\qquad
k_e(\phi,G)=\frac{C_m(\phi,G)}{C}.
\label{eq:k_defs}
\end{equation}
Equation~\eqref{eq:omega_pm} provides two directly observable quantities in both simulation and experiment: the twist-dependent dispersion of the hybridized branches $f_\pm(\phi,G)$ and their normal-mode splitting $\Omega(\phi,G)=|f_+-f_-|$. In driven one-port measurements, the visibility of both branches additionally depends on excitation/readout symmetry and on how strongly the loop couples to the corresponding current superpositions; consequently, the \emph{resolved} splitting can be suppressed near $\phi\simeq0$ and $\phi\simeq\pi/2$ even when near-field interaction is present. To capture this experimentally relevant resolvability in a compact way, we use the phenomenological envelope
\begin{equation}
\Omega_{\mathrm{res}}(\phi,G)\simeq \Omega_0(G)\,|\sin(2\phi)|,
\label{eq:Omega_envelope}
\end{equation}
which enforces suppression at the symmetry endpoints and maximal resolvability at intermediate angles.

\begin{figure}[!t]
\centering
\includegraphics[width=0.9\linewidth]{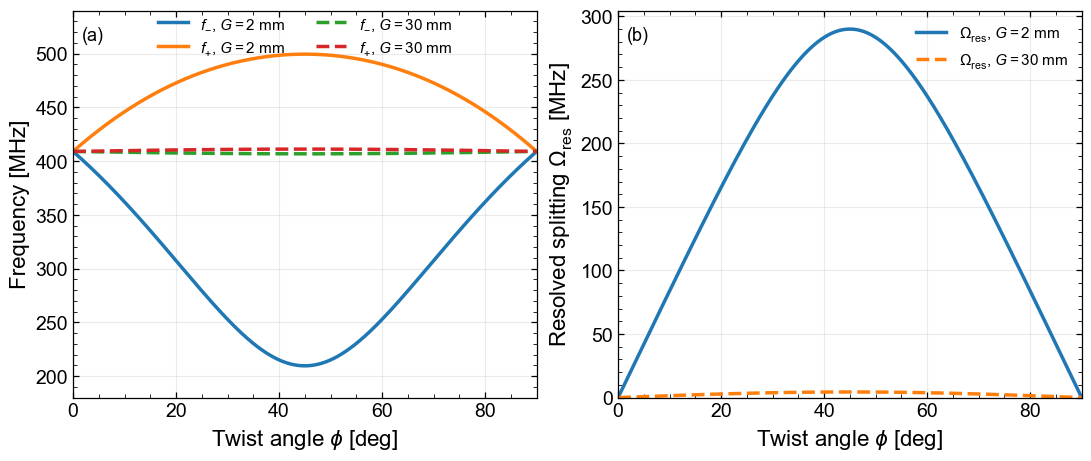}
\caption{Twist-controlled hybridization using the two-resonator model (Eqs.~\eqref{eq:omega_pm}--\eqref{eq:Omega_envelope}). (a) Predicted modal branches $f_{\pm}(\phi)$ for a strong-coupling gap ($G=2$~mm) and a weak-coupling gap ($G=30$~mm). (b) Corresponding resolved splitting $\Omega_{\mathrm{res}}(\phi)$. Parameters are chosen for the present structure scale ($L=350$~mm, $a\approx 5$~mm) to reproduce the observed order-of-magnitude tuning and to highlight the collapse of twist tunability with increased separation.}
\label{fig:theory_minimal}
\end{figure}

The qualitative implications of Eqs.~\eqref{eq:omega_pm}--\eqref{eq:Omega_envelope} are summarized in Fig.~\ref{fig:theory_minimal}. For a small gap (strong coupling), Fig.~\ref{fig:theory_minimal}(a) predicts a pronounced, continuous redshift of the lower branch $f_-(\phi)$ and a complementary blueshift of $f_+(\phi)$ as the effective coupling increases away from the symmetry endpoints, producing a large splitting that is maximal near intermediate angles. In contrast, for a large gap (weak coupling), both branches collapse toward an almost angle-independent reference frequency, and the splitting becomes too small to be readily resolved in a driven spectrum. Figure~\ref{fig:theory_minimal}(b) further emphasizes the distinction between physical interaction and experimental observability: even when coupling exists, the \emph{resolved} splitting is expected to be suppressed near $\phi\approx0$ and $\phi\approx\pi/2$ by symmetry/selection rules of the excitation and readout, and to peak at intermediate twist angles. This theory-only view provides a compact roadmap for interpreting the full-wave results in Fig.~\ref{fig:rotation_spectra_fields} and the one-port measurements reported later: strong, twist-programmable dispersion should appear only in the small-$G$ regime, and the most clearly resolvable hybridization should occur at intermediate $\phi$.

Because Eq.~\eqref{eq:omega_pm} involves two independent couplings, it can be inverted to separate inductive and capacitive contributions once $f_\pm(\phi,G)$ are extracted. Defining $A=(f_+/f_0)^2$ and $B=(f_-/f_0)^2$ with $f_0$ taken as an uncoupled reference (single-layer frequency or large-gap limit), one obtains
\begin{equation}
k_e(\phi,G)=\frac{2-A-B}{A-B},
\qquad
k_m(\phi,G)=A\big(1+k_e(\phi,G)\big)-1.
\label{eq:km_ke_inversion}
\end{equation}
This inversion is used later, together with linewidth extraction from the complex $S_{11}$ response \cite{probst2015robust,khalil2012analysis}, to report angle- and gap-dependent coupling and loss metrics. We define the normalized coupling and cooperativity as
\begin{equation}
g=\frac{\Omega}{2f_0},\qquad \Omega=|f_+-f_-|,
\qquad
\mathcal{C}\approx \frac{\Omega^{2}}{\Delta f_+\,\Delta f_-},
\label{eq:g_def}
\end{equation}
and use a purely classical resolved-splitting criterion, $\Omega \gtrsim (\Delta f_+ + \Delta f_-)/2$, to identify regimes where hybridization is spectrally resolvable in the driven response.

Finally, the gap dependence provides a direct diagnostic of the near-field origin of the tuning. As an order-of-magnitude guide, the per-unit-length parameters of two parallel cylindrical conductors scale with $\operatorname{acosh}(\cdot)$, illustrating that both inductive and capacitive interactions change rapidly with separation \cite{orfanidis2004ewa}. More generally, because the interaction is near-field dominated, the effective coupling decreases quickly with $G$; we parameterize this decay as
\begin{equation}
\kappa(G)\approx \kappa_0\,e^{-G/\xi},
\label{eq:kappa_gap}
\end{equation}
where $\xi$ is an effective near-field decay length set by the dominant transverse spatial content of the coupled bilayer. The comparison between small- and large-gap configurations in Secs.~B--D then serves as a clear experimental check that twist tunability is mediated by near-field coupling.

Guided by the minimal two-resonator picture summarized in Fig.~\ref{fig:theory_minimal}(a,b), we use full-wave simulations to (i) visualize how twist reshapes the near fields of the coupled bilayer and (ii) generate loop-driven $S_{11}$ traces under the same excitation geometry as the experiment, enabling a like-for-like comparison of twist trends. All simulations are performed in CST Microwave Studio (frequency-domain solver) with open boundaries and sufficient air padding around the structure; the wire grids and spacer follow the dimensions in Fig.~\ref{fig:dlwm_schematic}(a), and the loop-driven workflow follows the measurement concept in Fig.~\ref{fig:dlwm_schematic}(b).

Plane-wave simulations directly reveal the twist-controlled hybridization and the emergence of moir\'e-like magnetic near-field textures. Figure~\ref{fig:rotation_spectra_fields}(a) plots the peak mid-gap magnetic-field enhancement versus frequency for representative twist angles at small separation ($G=2$~mm). Consistent with Fig.~\ref{fig:theory_minimal}(a), increasing $\phi$ reshapes the spectrum in a manner characteristic of strong bilayer coupling: spectral features associated with the dominant manifold move and split, and new low-frequency features appear as hybridization becomes pronounced at intermediate angles. The field maps clarify the physical origin of this evolution. At $\phi=0^\circ$, the magnetic enhancement concentrates in the inter-wire regions while the electric enhancement remains largely tip-localized (Fig.~\ref{fig:rotation_spectra_fields}(b,c)), consistent with a standing-wave current that peaks near wire midpoints and charges that accumulate near wire ends. As $\phi$ increases, the mid-gap magnetic field develops stripe-like super-modulations whose orientation and spacing depend on twist (Fig.~\ref{fig:rotation_spectra_fields}(b)), whereas the electric field remains comparatively localized near the wire ends (Fig.~\ref{fig:rotation_spectra_fields}(c)). This behavior supports the interpretation that twist primarily reprograms the magnetic near field through interlayer current hybridization, while end-capacitive charge localization is less sensitive to the extended moir\'e-like beating.

To quantify the dominant magnetic super-modulation length, we extract an electromagnetic superperiod $\Lambda_{\mathrm{EM}}$ from the simulated $F_H$ maps in Fig.~\ref{fig:rotation_spectra_fields}(b). In the strongly coupled regime, we consistently find that the observable magnetic super-modulation is set by the hybridized near-field content of the bilayer and can substantially exceed $\Lambda_{\mathrm{geo}}$, emphasizing that the relevant ``moir\'e-like'' near-field texture is electromagnetic rather than purely geometric.

\begin{figure}[!t]
\centering
\includegraphics[width=0.8\linewidth]{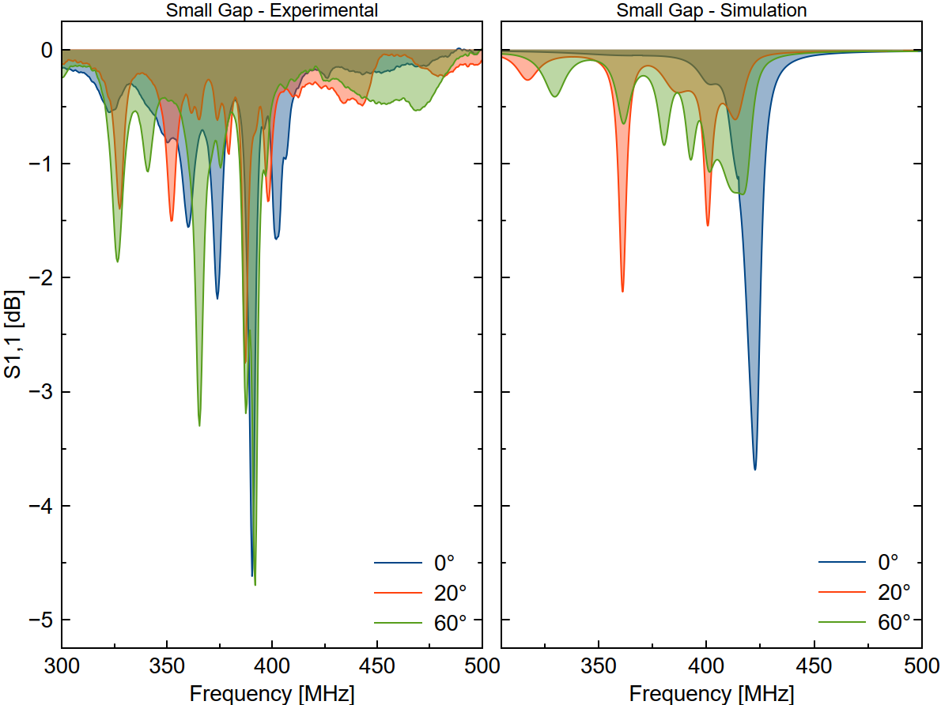}
\caption{One-port characterization in the strongly coupled (small-gap) regime, $G\approx3$~mm. (a) Measured referenced $S_{11}$ for representative twist angles. (b) Loop-driven full-wave simulations (emulating the probe geometry in Fig.~\ref{fig:dlwm_schematic}(b)) reproduce the main twist-dependent spectral trends; residual fine-structure differences are attributed to probe position/height tolerances and environment-dependent interference pathways.}
\label{fig:s11_exp_sim}
\end{figure}

For direct comparison with measurement, we also perform loop-driven (port) simulations that emulate the one-port excitation in Fig.~\ref{fig:dlwm_schematic}(b) and generate simulated $S_{11}(f,\phi,G)$ traces under the same excitation/readout symmetry. Figure~\ref{fig:s11_exp_sim}(a,b) compares measured and simulated $S_{11}$ spectra in the small-gap regime ($G\approx3$~mm). The experimental traces exhibit sharp angle-dependent notches and fine structure (Fig.~\ref{fig:s11_exp_sim}(a)), while the loop-driven simulations reproduce the principal twist trends and the locations of the dominant resonant features (Fig.~\ref{fig:s11_exp_sim}(b)). In both experiment and simulation, the strongest and most clearly resolved hybridization appears at intermediate $\phi$, consistent with the resolvability envelope anticipated by Fig.~\ref{fig:theory_minimal}(b): the bilayer interaction is present for small $G$, but the driven visibility of two distinct hybrid features is reduced near the symmetry endpoints and maximized away from them.

\begin{figure}[!t]
\centering
\includegraphics[width=0.8\linewidth]{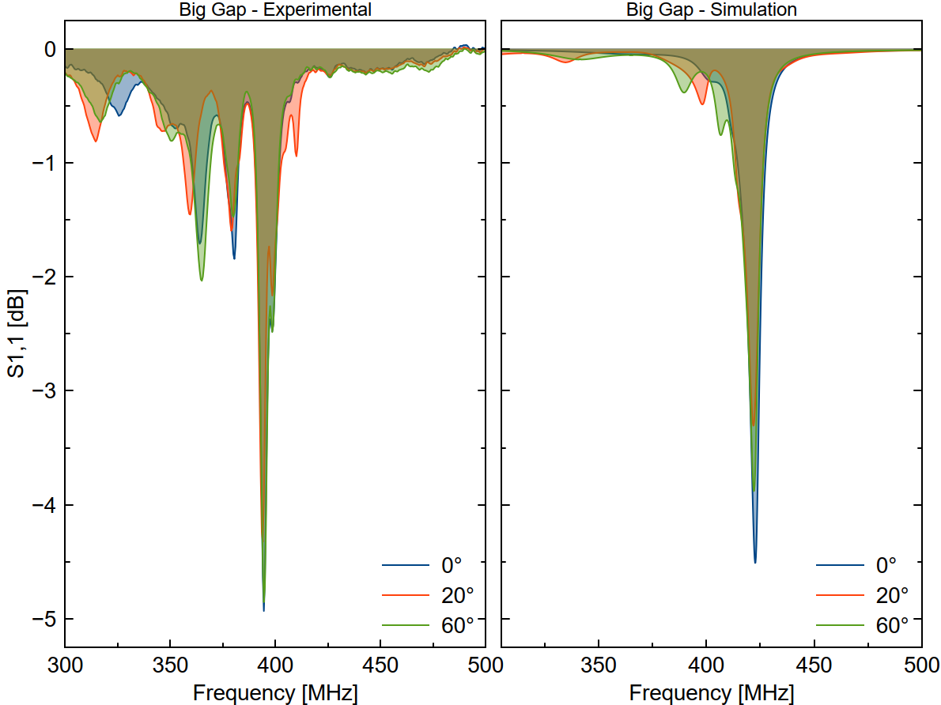}
\caption{Suppression of twist tunability in the weak-coupling (large-gap) regime, $G\approx30$~mm. (a) Measured referenced $S_{11}$ shows that the dominant resonance is largely insensitive to twist. (b) Loop-driven simulations reproduce the same collapse of twist tunability at large separation, confirming the near-field origin of the interaction.}
\label{fig:s11_large_gap}
\end{figure}

The gap dependence provides a second, decisive check of the near-field origin of twist tunability. Figure~\ref{fig:s11_large_gap}(a) shows that for a large separation ($G\approx30$~mm) the dominant measured resonance becomes largely insensitive to $\phi$, while Fig.~\ref{fig:s11_large_gap}(b) shows the same suppression of twist tunability in loop-driven simulations. This collapse of angle dispersion with increasing $G$ matches the qualitative trend in Fig.~\ref{fig:theory_minimal}(a,b) and is consistent with a rapidly decaying near-field interaction between the two grids. In this weak-coupling regime, the bilayer approaches two nearly independent sheets, so rotation predominantly alters coupling efficiency rather than shifting resonance frequencies.

\begin{figure}[!t]
\centering
\includegraphics[width=0.8\linewidth]{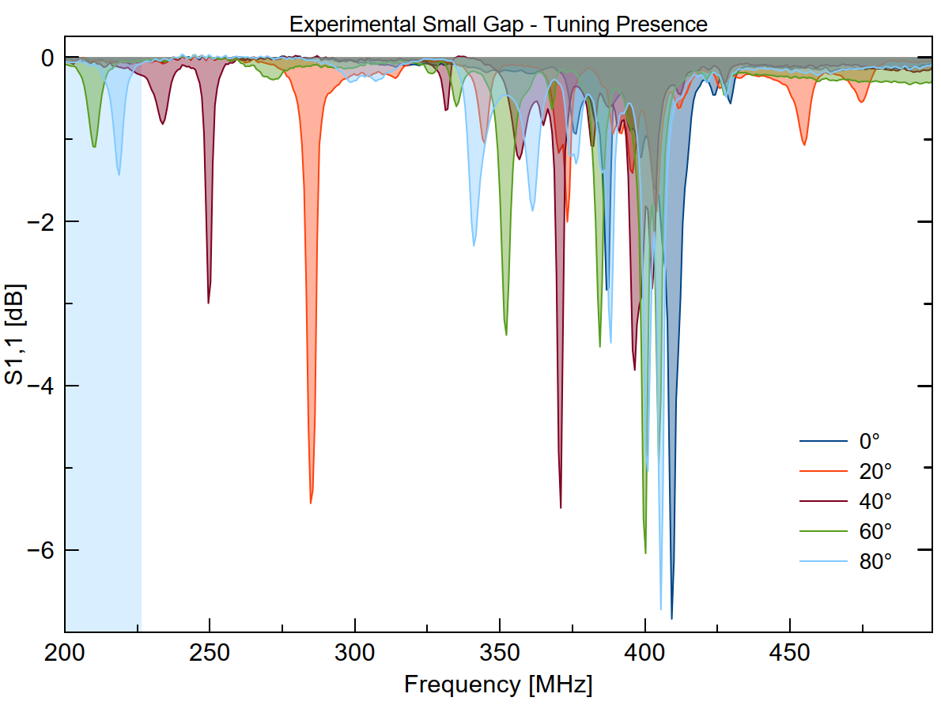}
\caption{Broadband evidence of twist-induced tuning in the strongly coupled (small-gap) regime. Referenced experimental $S_{11}$ over 200--500~MHz for multiple twist angles shows the appearance and continuous redshift of a low-frequency resonant feature into the $\sim$200--250~MHz band as $\phi$ increases, demonstrating large passive frequency control enabled solely by rotation.}
\label{fig:tuning_cmt}
\end{figure}

To capture the full extent of twist control in the strongly coupled configuration, Fig.~\ref{fig:tuning_cmt} presents broadband small-gap measurements over 200--500~MHz for multiple twist angles. The spectra show a clear, twist-dependent emergence and migration of a low-frequency resonant feature into the $\sim$200--250~MHz band, corroborating the large continuous redshift implied by the strong-coupling branch behavior in Fig.~\ref{fig:theory_minimal}(a). Using the electrical-size metric
\begin{equation}
\rho(\phi)=\frac{2L}{\lambda_{\mathrm{res}}(\phi)}=\frac{2L f_{\mathrm{res}}(\phi)}{c},
\label{eq:rho}
\end{equation}
the lowest observed resonance near $\sim$210~MHz for $L=350$~mm corresponds to $\rho\simeq 0.49$, demonstrating substantial resonant miniaturization enabled purely by twist-controlled bilayer coupling. Throughout, we avoid over-interpreting probe-sensitive fine structure in $S_{11}$ and instead base quantitative coupling metrics on resonance frequencies, linewidths, and splittings extracted from the complex response using the definitions established above. In particular, the strongest-coupled angles yield large normalized splitting and cooperativity consistent with the resolved-splitting (classical strong-coupling) regime, while the large-gap data in Fig.~\ref{fig:s11_large_gap}(a,b) provide the corresponding weak-coupling baseline.

\section{Discussion}
Twist-angle control has become a versatile knob in metasurfaces and metamaterials for shaping scattering, polarization, and dispersion, and more recently for moir\'e-enabled programmability and nonlinear responses \cite{chen2016review,holloway2012metasurfaces,wu2018moire,liu2022beamforming,liu2024reflective,liu2022magic,orazbay2024polar,valagiannopoulos2025nonlinear,dellavalle2017nonlinear}. The present work targets a different, antennas-and-propagation–motivated regime: a deeply subwavelength \emph{sub-GHz} resonant bilayer made of two \emph{identical} wire grids in which rotation is the \emph{only} in-situ control (Fig.~\ref{fig:dlwm_schematic}(a,b)). In the strong-coupling configuration (small $G$), both plane-wave field maps and loop-driven spectra show twist-controlled hybridization of the dominant resonant response: the simulated enhancement spectra evolve strongly with $\phi$ (Fig.~\ref{fig:rotation_spectra_fields}(a)), and the measured/loop-emulated one-port responses exhibit angle-dependent splitting and a pronounced low-frequency branch (Fig.~\ref{fig:s11_exp_sim}(a,b) and Fig.~\ref{fig:tuning_cmt}), consistent with the qualitative roadmap of the minimal two-resonator model (Fig.~\ref{fig:theory_minimal}(a,b)). The key practical consequence is large passive miniaturization: the lowest tracked resonance shifts from the $\sim$400~MHz range down to $\sim$210~MHz in the small-gap case, corresponding to an electrical size of $\rho\simeq 2Lf_{\mathrm{res}}/c\approx 0.49$ for $L=350$~mm (Eq.~\eqref{eq:rho}). The near-field origin of this effect is confirmed by the large-gap baseline: when $G$ is increased to 30~mm the twist dispersion collapses and the dominant resonance becomes nearly angle-insensitive in both experiment and simulation (Fig.~\ref{fig:s11_large_gap}(a,b)), as anticipated by the weak-coupling behavior in Fig.~\ref{fig:theory_minimal}(a,b). Beyond spectral tuning, twisting imprints programmable magnetic near-field super-modulations: the mid-gap magnetic enhancement develops stripe-like beating patterns whose orientation and spacing depend on $\phi$ (Fig.~\ref{fig:rotation_spectra_fields}(b)), while the electric field remains comparatively end-localized (Fig.~\ref{fig:rotation_spectra_fields}(c)). Importantly, the dominant superperiod is \emph{electromagnetic} rather than purely geometric: we extract $\Lambda_{\mathrm{EM}}$ directly from the field maps (via autocorrelation/line-cut–based spacing) and compare it to the geometric overlay scale $\Lambda_{\mathrm{geo}}=a/[2|\sin(\phi/2)|]$ (Eq.~\eqref{eq:Lambda_geo}); in the strongly coupled regime, $\Lambda_{\mathrm{EM}}$ can substantially exceed $\Lambda_{\mathrm{geo}}$, consistent with the broader understanding that moir\'e metasurface responses are governed by coupled-field overlap and spectral content rather than geometry alone \cite{wu2018moire,liu2022beamforming,liu2024reflective,mcguyer2022moire}. Finally, the ``busy'' appearance of raw one-port $S_{11}$ traces is expected for a near-field probe in the presence of a direct/background path and multiple resonant channels, which generically produces Fano-like line shapes and systematic fitting sensitivities \cite{fano1961,miroshnichenko2010fano,fan2003tcmt,rieger2023fano}. For this reason, coupling metrics reported here rely on resonance frequencies, linewidths, and splittings extracted from the \emph{complex} response using robust circle-fit–type procedures \cite{probst2015robust,khalil2012analysis,baity2024circlefit}. Within this strictly classical coupled-resonator interpretation, the small-gap data support large normalized splitting (up to $g\approx 0.43$) and cooperativity exceeding unity over broad angular ranges, while the large-gap case provides a clear weak-coupling control. The DLWM is intentionally macroscale and finite, so multimode/edge families and probe/environment sensitivity remain practical limitations; nevertheless, the combined agreement between the theory roadmap (Fig.~\ref{fig:theory_minimal}), full-wave field/spectrum trends (Fig.~\ref{fig:rotation_spectra_fields}), and one-port measurements (Figs.~\ref{fig:s11_exp_sim}--\ref{fig:tuning_cmt}) establishes twist as a simple, bias-free mechanism for programming sub-GHz resonant spectra and magnetic near fields, with straightforward extensions to printed implementations, two-port filtering, and multilayer programmable stacks.

\section{Conclusion}
We demonstrated a twist-controlled double-layer wire metasurface in the sub-GHz (VHF/UHF) band in which the relative in-plane rotation $\phi$ is the only tuning knob. In the strong near-field coupling regime (small $G$), twist drives robust hybridization and large passive tuning, including a continuous redshift of the lowest tracked resonance to $\sim$210~MHz, yielding an electrical size $\rho\simeq 0.49$ for $L=350$~mm. Full-wave field maps show that twist simultaneously programs moir\'e-like magnetic super-modulations with an electromagnetic superperiod $\Lambda_{\mathrm{EM}}$ extracted directly from the near-field patterns and generally distinct from the geometric overlay scale. Probe-aware analysis of complex one-port data yields large classical coupling figures of merit (normalized splitting up to $g\approx 0.43$ and cooperativity $>1$), while the collapse of tunability at large separation confirms the near-field origin of the interaction. Overall, the DLWM provides a compact route to bias-free, twist-programmable sub-GHz resonators/filters and magnetic near-field shapers that complements prior twist-angle metasurface work focused primarily on wavefront/polarization control or nonlinear switching.

\section*{Acknowledgments}
The author acknowledges financial support from the U.S. Department of Energy (DoE) and the U.S. Air Force Office of Scientific Research (AFOSR).

\section*{Competing interests}
The authors declare no competing interests.

\bibliographystyle{unsrt}
\bibliography{references}

\end{document}